\documentclass[12]{article}
\usepackage{fleqn}
\usepackage{graphicx}
\begin{document}
\Large
\begin{center}
{\bf Projective Ring Line Encompassing\\ Two-Qubits}
\end{center}
\vspace*{.2cm}
\begin{center}
Metod Saniga,$^{1}$ Michel Planat$^{2}$
and Petr Pracna$^{3}$
\end{center}
\vspace*{.0cm} \normalsize
\begin{center}
$^{1}$Astronomical Institute, Slovak Academy of Sciences\\
SK-05960 Tatransk\' a Lomnica, Slovak Republic\\
(msaniga@astro.sk)

\vspace*{.4cm} $^{2}$Institut FEMTO-ST, CNRS, D\' epartement LPMO,
32 Avenue de
l'Observatoire\\ F-25044 Besan\c con Cedex, France\\
(michel.planat@femto-st.fr)

\vspace*{.4cm} $^{3}$J. Heyrovsk\' y Institute of Physical
Chemistry, Academy of Sciences of the Czech Republic\\ Dolej\v
skova 3, CZ-182 23 Prague 8, Czech
Republic\\
(petr.pracna@jh-inst.cas.cz)

\end{center}

\vspace*{.2cm} \noindent \hrulefill

\vspace*{.1cm} \noindent {\bf Abstract}

\noindent The projective line over the (non-commutative) ring of
two-by-two matrices with coefficients in $GF(2)$ is found to {\it
fully} accommodate the algebra of 15 operators --- generalized
Pauli matrices --- characterizing two-qubit systems. The relevant
sub-configuration consists of 15 points each of which is either
simultaneously distant or simultaneously neighbor to (any) two
given distant points of the line. The operators can be identified
with the points in such a one-to-one manner that their commutation
relations are {\it exactly} reproduced by the underlying geometry
of the points, with the ring geometrical notions of
neighbor/distant answering, respectively, to the operational ones
of commuting/non-commuting. This remarkable configuration can be
viewed in two principally different ways accounting, respectively,
for the basic 9+6 and 10+5 factorizations of the algebra of the
observables. First, as a disjoint union of the projective line
over $GF(2)\times GF(2)$ (the ``Mermin" part) and two lines over
$GF(4)$ passing through the two selected points, the latter
omitted. Second, as the generalized quadrangle of order two, with its
ovoids and/or spreads standing for (maximum) sets of five mutually non-commuting
operators and/or groups of five maximally commuting subsets of three operators each.
These findings open up rather
unexpected vistas for an algebraic geometrical modelling of
finite-dimensional quantum systems and give their numerous
applications a wholly new perspective.

\vspace*{-.1cm} \noindent \hrulefill

\vspace*{.5cm}
\noindent
Projective lines defined over finite associative rings with
unity/identity$^{1-7}$ have recently been recognized to be an
important novel tool for getting a deeper insight into the
underlying algebraic geometrical structure of finite dimensional
quantum systems.$^{8-10}$ Focusing almost uniquely on the
two-qubit case, i.e., the set of 15 operators/generalized
four-by-four Pauli spin matrices, of particular importance turned
out to be the lines defined over the direct product of the
simplest Galois fields, $GF(2) \times GF(2) \times \ldots \times
GF(2)$. Here, the line defined over $GF(2) \times GF(2)$ plays a
prominent role in grasping qualitatively the basic structure of
so-called Mermin squares,$^{9,10}$ i.\,e., three-by-three arrays
in certain remarkable 9 + 6 split-ups of the algebra of operators,
whereas the line over  $GF(2) \times GF(2) \times GF(2)$ reflects
some of the basic features of a specific 8 + 7
(``cube-and-kernel") factorization of the set.$^{10}$  Motivated
by these partial findings, we started our quest for such a ring
line that would provide us with a complete picture of the algebra
of all the 15 operators/matrices. After examining a large number
of lines defined over commutative rings,$^{6,7}$ we gradually
realized that a proper candidate is likely to be found in the
non-commutative domain and this, indeed, turned out to be a right
move. It is, as we shall demonstrate in sufficient detail, the
projective line defined  over the full two-by-two matrix ring with
entries in $GF(2)$ --- the unique simple non-commutative ring of
order 16 featuring six units (invertible elements) and ten
zero-divisors.$^{11}$ Having in mind the conceptual rather than
formal side of the task, we shall try to reduce the technicalities
of the exposition to a minimum, referring instead the interested
reader to the relevant literature.

We first recall the concept of a projective ring line.$^{1-7}$
Given an associative ring $R$ with unity/identity$^{12-14}$ and $GL(2,R)$, the
general linear group of invertible two-by-two matrices with
entries in $R$, a pair $(a, b) \in R^{2}$ is called admissible
over $R$ if there exist $c, d \in R$ such that
\begin{equation}
\left(
\begin{array}{cc}
a & b \\
c & d \\
\end{array}
\right) \in GL(2,R).
\end{equation}
The projective line over $R$, usually denoted as $P_{1}(R)$, is
the set of equivalence classes of ordered pairs $(\varrho a,
\varrho b)$, where $\varrho$ is a unit of $R$ and $(a, b)$ is
admissible. Two points $X : =  (\varrho a, \varrho b)$ and $Y : =
(\varrho c, \varrho d)$ of the line are called {\it distant} or
{\it neighbor} according as
\begin{equation}
\left(
\begin{array}{cc}
a & b \\
c & d \\
\end{array}
\right) \in GL(2,R)~~~~{\rm or}~~~~\left(
\begin{array}{cc}
a & b \\
c & d \\
\end{array}
\right) \notin GL(2,R),
\end{equation}
respectively. $GL(2,R)$ has an important property of acting transitively on a set of three pairwise distant points;
that is, given any two triples of mutually distant points there exists an element of $GL(2,R)$ transforming one triple
into the other.

     The projective line we are exclusively interested in here is the one defined over the full two-by-two matrix ring
     with GF(2)-valued coefficients, i.\,e.,
\begin{equation}
    R = M_{2}(GF(2)) \equiv \left\{ \left(
\begin{array}{cc}
\alpha & \beta \\
\gamma & \delta \\
\end{array}
\right) \mid ~ \alpha, \beta, \gamma, \delta \in GF(2) \right\}.
\end{equation}
Labelling these matrices as follows
\begin{eqnarray}
&&~1 \equiv \left(
\begin{array}{cc}
1 & 0 \\
0 & 1 \\
\end{array}
\right),~2 \equiv \left(
\begin{array}{cc}
0 & 1 \\
1 & 0 \\
\end{array}
\right),~ 3 \equiv \left(
\begin{array}{cc}
1 & 1 \\
1 & 1 \\
\end{array}
\right),~
4 \equiv \left(
\begin{array}{cc}
0 & 0 \\
1 & 1 \\
\end{array}
\right), \nonumber \\
&&~5 \equiv \left(
\begin{array}{cc}
1 & 0 \\
1 & 0 \\
\end{array}
\right),~6 \equiv \left(
\begin{array}{cc}
0 & 1 \\
0 & 1 \\
\end{array}
\right),~ 7 \equiv \left(
\begin{array}{cc}
1 & 1 \\
0 & 0 \\
\end{array}
\right),~
8 \equiv \left(
\begin{array}{cc}
0 & 1 \\
0 & 0 \\
\end{array}
\right), \nonumber \\
&&~9 \equiv \left(
\begin{array}{cc}
1 & 1 \\
0 & 1 \\
\end{array}
\right),~10 \equiv \left(
\begin{array}{cc}
0 & 0 \\
1 & 0 \\
\end{array}
\right),~11 \equiv \left(
\begin{array}{cc}
1 & 0 \\
1 & 1 \\
\end{array}
\right),~12 \equiv \left(
\begin{array}{cc}
0 & 1 \\
1 & 1 \\
\end{array}
\right), \nonumber \\
&&13 \equiv \left(
\begin{array}{cc}
1 & 1 \\
1 & 0 \\
\end{array}
\right),~14 \equiv \left(
\begin{array}{cc}
0 & 0 \\
0 & 1 \\
\end{array}
\right),~15 \equiv \left(
\begin{array}{cc}
1 & 0 \\
0 & 0 \\
\end{array}
\right),~0 \equiv \left(
\begin{array}{cc}
0 & 0 \\
0 & 0 \\
\end{array}
\right),
\end{eqnarray}
one can readily verify that addition and multiplication in
$M_{2}(GF(2))$ is carried out as shown in Table 1.$^{15}$ Checking
first for admissibility (Eq.\,(1)) and then grouping the
admissible pairs left-proportional by a unit into equivalence
classes (of cardinality six each), we find that
$P_{1}(M_{2}(GF(2)))$\footnote{This line has been found to have a
distinguished footing among non-commutative ring lines for it
fundamentally differs from its two commutative
counterparts.$^{11}$} possesses altogether 35 points, with the
following representatives of each equivalence class (see
Refs.\,6--8 for more details about this methodology and  a number
of illustrative examples of a projective ring line):
\begin{eqnarray}
&&(1,1),~(1,2),~(1,9),~(1,11),~(1,12), (1,13), \nonumber \\
&&(1,0),~(1,3),~(1,4),~(1,5),~(1,6),~(1,7),~(1,8),~(1,10),~(1,14),~(1,15), \nonumber \\
&&(0,1),~(3,1),~(4,1),~(5,1),~(6,1),~(7,1),~(8,1),~(10,1),~(14,1),~(15,1), \nonumber \\
&&(3,4),~(3,10),~(3,14),~(5,4),~(5,10),~(5,14),~(6,4),~(6,10),~(6,14).
\end{eqnarray}
From the multiplication table one can easily recognize that the representatives in the first row of the last equation
have both entries units (1 being, obviously, unity/multiplicative identity), those of the second and third row have
one entry unit(y) and the other a zero-divisor, whilst all pairs in the last row feature zero-divisors in both the entries.
At this point we are ready to shown which ``portion" of $P_{1}(M_{2}(GF(2)))$ is the proper algebraic geometrical setting
of two-qubits.

\begin{table}[t]
\begin{center}
\caption{Addition ({\it top}) and multiplication ({\it bottom}) in
$M_{2}(GF(2))$.} \vspace*{0.2cm}
\begin{tabular}{||c|cccccccccccccccc||}
\hline \hline
$+$ & $0$ & $1$ & $2$ & $3$ & $4$ & $5$ & $6$ & $7$ & $8$ & $9$ & $10$ & $11$ & $12$ & $13$ & $14$ & $15$\\
\hline
$0$ & $0$ & $1$ & $2$ & $3$ & $4$ & $5$ & $6$ & $7$ & $8$ & $9$ & $10$ & $11$ & $12$ & $13$ & $14$ & $15$ \\
$1$ & $1$ & $0$ & $3$ & $2$ & $5$ & $4$ & $7$ & $6$ & $9$ & $8$ & $11$ & $10$ & $13$ & $12$ & $15$ & $14$\\
$2$ & $2$ & $3$ & $0$ & $1$ & $6$ & $7$ & $4$ & $5$ & $10$ & $11$ & $8$ & $9$ & $14$ & $15$ & $12$ & $13$\\
$3$ & $3$ & $2$ & $1$ & $0$ & $7$ & $6$ & $5$ & $4$ & $11$ & $10$ & $9$ & $8$ & $15$ & $14$ & $13$ & $12$\\
$4$ & $4$ & $5$ & $6$ & $7$ & $0$ & $1$ & $2$ & $3$ & $12$ & $13$ & $14$ & $15$ & $8$ & $9$ & $10$ & $11$\\
$5$ & $5$ & $4$ & $7$ & $6$ & $1$ & $0$ & $3$ & $2$ & $13$ & $12$ & $15$ & $14$ & $9$ & $8$ & $11$ & $10$\\
$6$ & $6$ & $7$ & $4$ & $5$ & $2$ & $3$ & $0$ & $1$ & $14$ & $15$ & $12$ & $13$ & $10$ & $11$ & $8$ & $9$\\
$7$ & $7$ & $6$ & $5$ & $4$ & $3$ & $2$ & $1$ & $0$ & $15$ & $14$ & $13$ & $12$ & $11$ & $10$ & $9$ & $8$\\
$8$ & $8$ & $9$ & $10$ & $11$ & $12$ & $13$ & $14$ & $15$ & $0$ & $1$ & $2$ & $3$ & $4$ & $5$ & $6$ & $7$\\
$9$ & $9$ & $8$ & $11$ & $10$ & $13$ & $12$ & $15$ & $14$ & $1$ & $0$ & $3$ & $2$ & $5$ & $4$ & $7$ & $6$\\
$10$ & $10$ & $11$ & $8$ & $9$ & $14$ & $15$ & $12$ & $13$ & $2$ & $3$ & $0$ & $1$ & $6$ & $7$ & $4$ & $5$\\
$11$ & $11$ & $10$ & $9$ & $8$ & $15$ & $14$ & $13$ & $12$ & $3$ & $2$ & $1$ & $0$ & $7$ & $6$ & $5$ & $4$\\
$12$ & $12$ & $13$ & $14$ & $15$ & $8$ & $9$ & $10$ & $11$ & $4$ & $5$ & $6$ & $7$ & $0$ & $1$ & $2$ & $3$\\
$13$ & $13$ & $12$ & $15$ & $14$ & $9$ & $8$ & $11$ & $10$ & $5$ & $4$ & $7$ & $6$ & $1$ & $0$ & $3$ & $2$\\
$14$ & $14$ & $15$ & $12$ & $13$ & $10$ & $11$ & $8$ & $9$ & $6$ & $7$ & $4$ & $5$ & $2$ & $3$ & $0$ & $1$\\
$15$ & $15$ & $14$ & $13$ & $12$ & $11$ & $10$ & $9$ & $8$ & $7$ & $6$ & $5$ & $4$ & $3$ & $2$ & $1$ & $0$\\
\hline \hline
&&&&&&&&&&&&&&&&\\
\hline \hline
$\times$ &
      $0$ & $1$ & $2$ & $3$ & $4$ & $5$ & $6$ & $7$ & $8$ & $9$ & $10$ & $11$ & $12$ & $13$ & $14$ & $15$\\
\hline
$0$ & $0$ & $0$ & $0$ & $0$ & $0$ & $0$ & $0$ & $0$ & $0$ & $0$ & $0$ & $0$ & $0$ & $0$ & $0$ & $0$ \\
$1$ & $0$ & $1$ & $2$ & $3$ & $4$ & $5$ & $6$ & $7$ & $8$ & $9$ & $10$ & $11$ & $12$ & $13$ & $14$ & $15$\\
$2$ & $0$ & $2$ & $1$ & $3$ & $7$ & $5$ & $6$ & $4$ & $14$ & $12$ & $15$ & $13$ & $9$ & $11$ & $8$ & $10$\\
$3$ & $0$ & $3$ & $3$ & $0$ & $3$ & $0$ & $0$ & $3$ & $6$ & $5$ & $5$ & $6$ & $5$ & $6$ & $6$ & $5$\\
$4$ & $0$ & $4$ & $4$ & $0$ & $4$ & $0$ & $0$ & $4$ & $14$ & $10$ & $10$ & $14$ & $10$ & $14$ & $14$ & $10$\\
$5$ & $0$ & $5$ & $6$ & $3$ & $0$ & $5$ & $6$ & $3$ & $6$ & $3$ & $0$ & $5$ & $6$ & $3$ & $0$ & $5$\\
$6$ & $0$ & $6$ & $5$ & $3$ & $3$ & $5$ & $6$ & $0$ & $0$ & $6$ & $5$ & $3$ & $3$ & $5$ & $6$ & $0$\\
$7$ & $0$ & $7$ & $7$ & $0$ & $7$ & $0$ & $0$ & $7$ & $8$ & $15$ & $15$ & $8$ & $15$ & $8$ & $8$ & $15$\\
$8$ & $0$ & $8$ & $15$ & $7$ & $7$ & $15$ & $8$ & $0$ & $0$ & $8$ & $15$ & $7$ & $7$ & $15$ & $8$ & $0$\\
$9$ & $0$ & $9$ & $13$ & $4$ & $3$ & $10$ & $14$ & $7$ & $8$ & $1$ & $5$ & $12$ & $11$ & $2$ & $6$ & $15$\\
$10$ & $0$ & $10$ & $14$ & $4$ & $0$ & $10$ & $14$ & $4$ & $14$ & $4$ & $0$ & $10$ & $14$ & $4$ & $0$ & $10$\\
$11$ & $0$ & $11$ & $12$ & $7$ & $4$ & $15$ & $8$ & $3$ & $6$ & $13$ & $10$ & $1$ & $2$ & $9$ & $14$ & $5$\\
$12$ & $0$ & $12$ & $11$ & $7$ & $3$ & $15$ & $8$ & $4$ & $14$ & $2$ & $5$ & $9$ & $13$ & $1$ & $6$ & $10$\\
$13$ & $0$ & $13$ & $9$ & $4$ & $7$ & $10$ & $14$ & $3$ & $6$ & $11$ & $15$ & $2$ & $1$ & $12$ & $8$ & $5$\\
$14$ & $0$ & $14$ & $10$ & $4$ & $4$ & $10$ & $14$ & $0$ & $0$ & $14$ & $10$ & $4$ & $4$ & $10$ & $14$ & $0$\\
$15$ & $0$ & $15$ & $8$ & $7$ & $0$ & $15$ & $8$ & $7$ & $8$ & $7$ & $0$ & $15$ & $8$ & $7$ & $0$ & $15$\\
\hline \hline
\end{tabular}
\end{center}
\end{table}

To this end, we consider two distant points of the line. Taking
into account the above-mentioned three-distant-transitivity of
$GL(2,R)$, we can take these, without any loss of generality, to
be the points $U:= (1,0)$ and $V:=(0,1)$. Next we pick up all
those points of the line which are either simultaneously distant
or simultaneously neighbor to $U$ and $V$. Employing the left part
of Eq.\,(2), we find the following six points
\begin{eqnarray}
&&C_{1}=(1,1),~C_{2}=(1,2),~C_{3}=(1,9), \nonumber \\
&&C_{4}=(1,11),~C_{5}=(1,12),~C_{6}=(1,13),
\end{eqnarray}
to belong to the first family, whereas the right part of Eq.\,(2) tells us that the second family comprises
the following nine points
\begin{eqnarray}
&&C_{7}=(3,4),~C_{8}=(3,10),~C_{9}=(3,14), \nonumber \\
&&C_{10}=(5,4),~C_{11}=(5,10),~C_{12}=(5,14),  \nonumber \\
&&C_{13}=(6,4),~C_{14}=(6,10),~C_{15}=(6,14).
\end{eqnarray}
Making again use of Eq.\,(2), one finds that the points of our
special subset of  $P_{1}(M_{2}(GF(2)))$ are related with each
other as shown in Table 2; from this table it can readily be
discerned that to every point of the configuration there are six
neighbor and eight distant points, and that the maximum number of
pairwise neighbor points is three.
\begin{table}[t]
\begin{center}
\caption{The distant and neighbor (``+" and ``$-$", respectively)
relation between the points of the configuration. The points are
arranged in such a way that the last nine of them (i.\,e., $C_{7}$
to $C_{15}$) form the projective line over $GF(2) \times
GF(2)$.$^{8-10}$} \vspace*{0.4cm}
\begin{tabular}{||c|ccc|ccc|ccc|ccc|ccc||}
\hline \hline
 & $C_{1}$ & $C_{2}$ & $C_{3}$ & $C_{4}$ & $C_{5}$ & $C_{6}$ & $C_{7}$ & $C_{8}$ & $C_{9}$ & $C_{10}$
& $C_{11}$ & $C_{12}$ & $C_{13}$ & $C_{14}$ & $C_{15}$ \\
\hline
$C_{1}$ & $-$ & $-$ & $-$ & $-$ & $+$ & $+$ & $-$ & $+$ & $+$ & $+$ & $-$ & $+$ & $+$ & $+$ & $-$ \\
$C_{2}$ & $-$ & $-$ & $+$ & $+$ & $-$ & $-$ & $-$ & $+$ & $+$ & $+$ & $+$ & $-$ & $+$ & $-$ & $+$ \\
$C_{3}$ & $-$ & $+$ & $-$ & $+$ & $-$ & $-$ & $+$ & $-$ & $+$ & $-$ & $+$ & $+$ & $+$ & $+$ & $-$ \\
\hline
$C_{4}$ & $-$ & $+$ & $+$ & $-$ & $-$ & $-$ & $+$ & $+$ & $-$ & $+$ & $-$ & $+$ & $-$ & $+$ & $+$ \\
$C_{5}$ & $+$ & $-$ & $-$ & $-$ & $-$ & $+$ & $+$ & $-$ & $+$ & $+$ & $+$ & $-$ & $-$ & $+$ & $+$ \\
$C_{6}$ & $+$ & $-$ & $-$ & $-$ & $+$ & $-$ & $+$ & $+$ & $-$ & $-$ & $+$ & $+$ & $+$ & $-$ & $+$ \\
\hline
$C_{7}$ & $-$ & $-$ & $+$ & $+$ & $+$ & $+$ & $-$ & $-$ & $-$ & $-$ & $+$ & $+$ & $-$ & $+$ & $+$ \\
$C_{8}$ & $+$ & $+$ & $-$ & $+$ & $-$ & $+$ & $-$ & $-$ & $-$ & $+$ & $-$ & $+$ & $+$ & $-$ & $+$ \\
$C_{9}$ & $+$ & $+$ & $+$ & $-$ & $+$ & $-$ & $-$ & $-$ & $-$ & $+$ & $+$ & $-$ & $+$ & $+$ & $-$ \\
\hline
$C_{10}$ & $+$ & $+$ & $-$ & $+$ & $+$ & $-$ & $-$ & $+$ & $+$ & $-$ & $-$ & $-$ & $-$ & $+$ & $+$ \\
$C_{11}$ & $-$ & $+$ & $+$ & $-$ & $+$ & $+$ & $+$ & $-$ & $+$ & $-$ & $-$ & $-$ & $+$ & $-$ & $+$ \\
$C_{12}$ & $+$ & $-$ & $+$ & $+$ & $-$ & $+$ & $+$ & $+$ & $-$ & $-$ & $-$ & $-$ & $+$ & $+$ & $-$ \\
\hline
$C_{13}$ & $+$ & $+$ & $+$ & $-$ & $-$ & $+$ & $-$ & $+$ & $+$ & $-$ & $+$ & $+$ & $-$ & $-$ & $-$ \\
$C_{14}$ & $+$ & $-$ & $+$ & $+$ & $+$ & $-$ & $+$ & $-$ & $+$ & $+$ & $-$ & $+$ & $-$ & $-$ & $-$ \\
$C_{15}$ & $-$ & $+$ & $-$ & $+$ & $+$ & $+$ & $+$ & $+$ & $-$ & $+$ & $+$ & $-$ & $-$ & $-$ & $-$ \\
\hline \hline
\end{tabular}
\end{center}
\end{table}
The final step is to identify these 15 points with the 15 generalized Pauli matrices/operators of two-qubits (see, e.\,g.,
Ref.\,10, Eq.\,(1)) in the following way
\begin{eqnarray}
&&C_{1}= \sigma_z \otimes \sigma_x,~C_{2}= \sigma_y \otimes \sigma_y,~C_{3}= 1_2 \otimes \sigma_x, \nonumber \\
&&C_{4}= \sigma_y \otimes \sigma_z,~C_{5}= \sigma_y \otimes 1_2,~C_{6}= \sigma_x \otimes \sigma_x, \nonumber \\
&&C_{7}= \sigma_x \otimes \sigma_z,~C_{8}= \sigma_y \otimes \sigma_x,~C_{9}= \sigma_z \otimes \sigma_y, \nonumber \\
&&C_{10}= \sigma_x\otimes 1_2,~C_{11}= \sigma_x \otimes \sigma_y,~C_{12}= 1_2\otimes \sigma_y,  \nonumber \\
&&C_{13}= 1_2\otimes \sigma_z,~C_{14}= \sigma_z \otimes \sigma_z,~C_{15}= \sigma_z\otimes 1_2,
\end{eqnarray}
where $1_2$ is the $2 \times 2$ unit matrix, $\sigma_x$,
$\sigma_y$ and $\sigma_z$ are the classical Pauli matrices and the
symbol ``$\otimes$" stands for the tensorial product of matrices,
in order to readily verify that Table 2 gives the correct
commutation relations between these operators with the symbols
``+" and ``$-$" now having the meaning of ``non-commuting"  and
``commuting", respectively. Slightly rephrased, {\it one and the
same} ``incidence matrix", Table 2, pertains to {\it two distinct}
configurations of a {\it completely different} origin: a set of
points of the projective line over a particular finite ring, with
the symbols ``$+$"/``$-$"  having the algebraic geometrical meaning
of distant/neighbor, as well as a set of operators of
four-dimensional Hilbert space, with the same symbols acquiring
the operational meaning of non-commuting/commuting, respectively.

This remarkable configuration can be interpreted in two principally different ways, which account,
respectively, for the basic 9+6 (Fig.\,1, {\it left}) and 10+5 (Fig.\,1, {\it right}) factorizations of the
algebra of observables. The first is simply a disjoint union of the projective line over $GF(2)\times GF(2)$
and two lines over $GF(4)$ passing through the two selected points $U$ and $V$, these latter omitted.  As demonstrated in
detail elsewhere,$^{9,10}$ the line over  $GF(2)\times GF(2)$ underlies the qualitative structure of Mermin's
magic squares, i.\,e., $3 \times 3$ arrays of nine observables commuting pairwise in each row and column and
arranged so that their product properties contradict those of the assigned eigenvalues.   The two lines over
$GF(4)$ represent the remaining, bipartite part of the split-up, where three points/observables on each of the
lines are mutually distant/non-commuting and every point/observable of one line is neighbor to/commutes with
any point/observable of the other line (see Fig.\,1, {\it left}).
\begin{figure}[h]
\centerline{\includegraphics[width=13.0truecm,clip=]{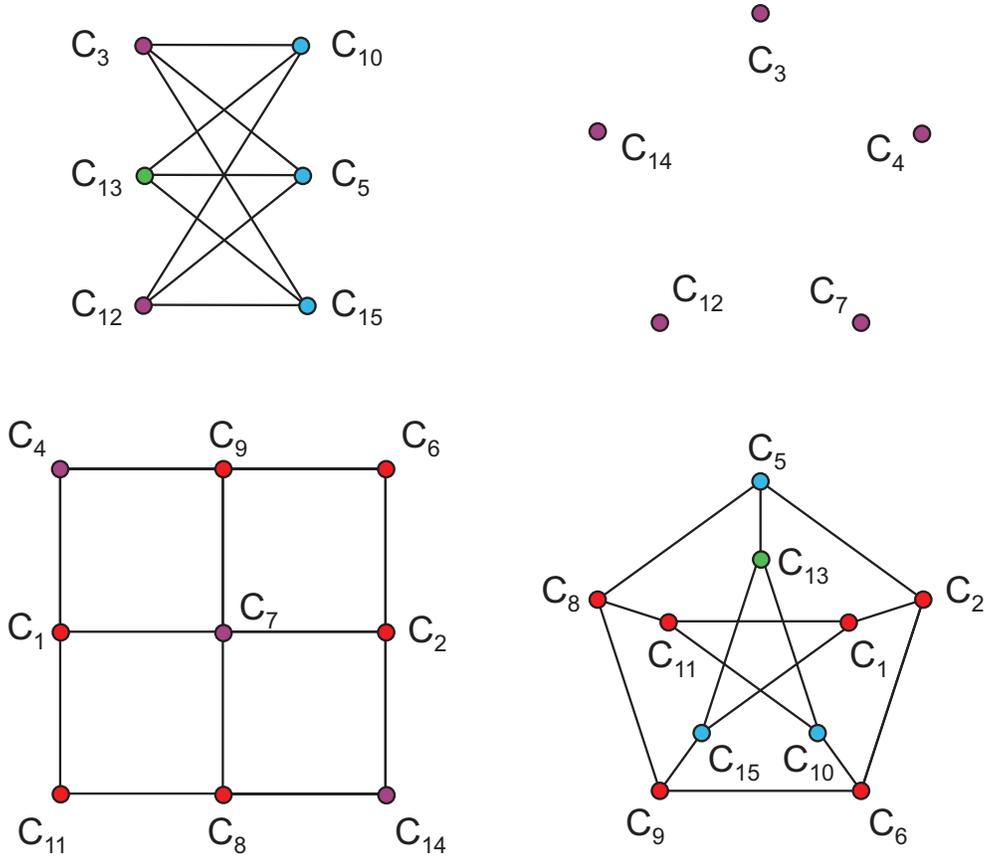}}
\vspace*{.5cm} \caption{The two basic factorizations of the
algebra of the 15 observables/operators of a two-qubit system. A
9+6 factorization ({\it left}) corresponds geometrically to the
split-up of our sub-configuration of $P_1(M_2(GF(2)))$ into the
projective line over $GF(2)\times GF(2)$ ({\it bottom}) and a
couple of projective lines over $GF(4)$ having two points in
common ({\it top}). A 10+5 one ({\it right}) corresponds, as also
demonstrated in a different way in Fig.\,2, to the partition of
the generalized quadrangle into one of its ovoids ({\it top}) and
the Petersen graph ({\it bottom}). In both the cases, two
points/observables are joined by a line-segment only if they are
neighbor/commute and the color is used to illustrate how the two
factorizations relate to each other.}
\end{figure}
The second interpretation involves a {\it generalized quadrangle},
a rank two point-line incidence geometry where two points share at
most one line and where for any point $X$ and a line ${\cal L}$,
$X \notin {\cal L}$, there exists exactly one line through $X$
which intersect ${\cal L}$.$^{16-20}$ The generalized quadrangle
associated with our observables is of order two, i.\,e., the one
where every line contains three points and every point is on three
lines. Such a quadrangle has, indeed, 15 points (and, because of
its self-duality, the same number of lines), each of which is
joined by a line with other six (Fig.\,2, {\it left}). If one
removes from this quadrangle one of its {\it ovoids}, i.\,e., a
set of (five) points that has exactly one point in common with
every line (Fig.\,2, {\it middle}), one is left with the set of
ten points that form the famous Petersen graph (Fig.\,2, {\it
right});$^{19,20}$ five points of an ovoid answer to nothing but
the five mutually distant points of $P_1(GF(4))$ and, so, to the
five (i.\,e., the maximum number of) mutually non-commuting
observables of two-qubits. If, dually, one removes from the
quadrangle a {\it spread}, i.\,e., a set of (five) pairwise
disjoint lines that partition the point set, one gets the dual of
the Petersen graph (Fig.\,3); five lines of a spread represent
nothing but the five maximum subsets of three mutually commuting
operators each, whose associated bases are {\it mutually
unbiased}.$^{10,21}$ It is a straightforward exercise to associate
the points of the quadrangle with the operators/observables $C_i$,
Eq.\,(8), in such a way to recover Table 2, after substituting the
``$-$"/``$+$" sign for any two points of the quadrangle which
are/are not on a common line.

\begin{figure}[t]
\centerline{\includegraphics[width=14.0truecm,clip=]{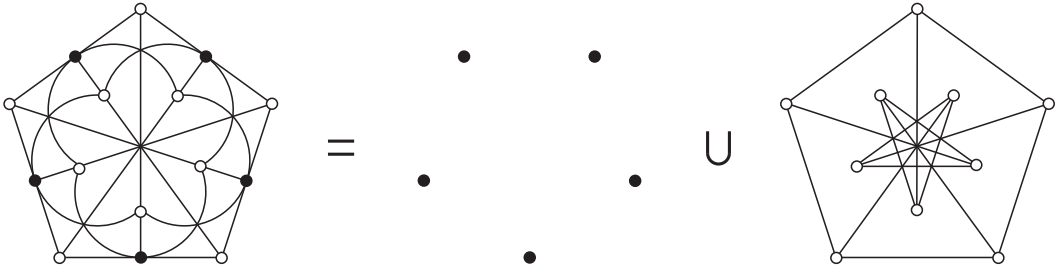}}
\vspace*{.2cm}
\caption{The generalized quadrangle of order two ({\it left}) and its factorization into an  ovoid  ({\it middle})
and the ``Petersen" part ({\it right}). The lines of the quadrangle are illustrated by the straight segments as
well as by the segments of circles; note that not every intersection of two segments counts for a point of
the quadrangle.}
\end{figure}
\begin{figure}[ht]
\centerline{\includegraphics[width=14.0truecm,clip=]{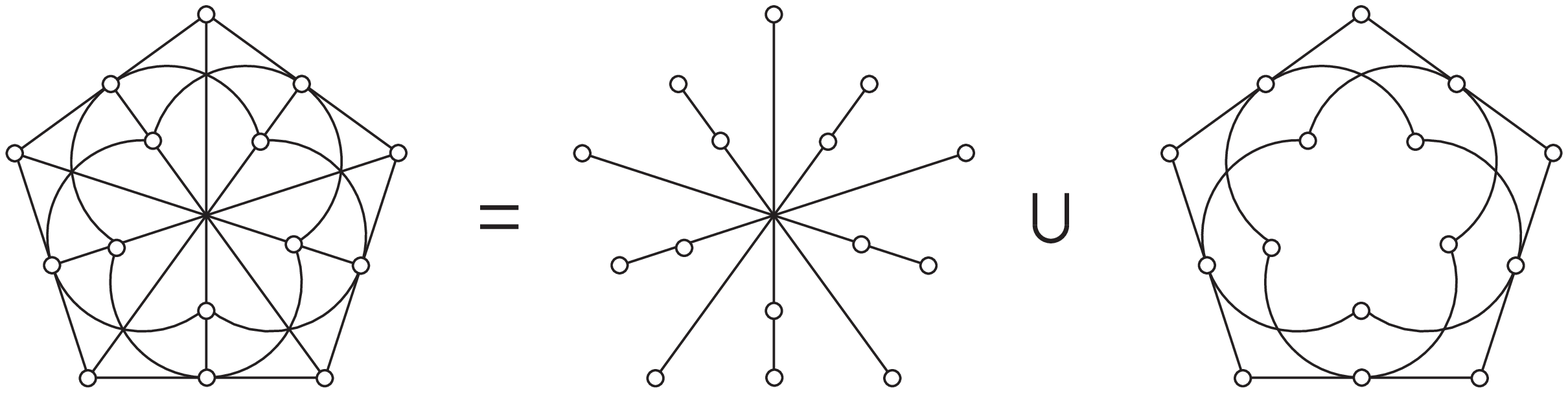}}
\vspace*{.2cm} \caption{A dual view of the generalized quadrangle
of order two ({\it left}) as a disjoint union of one of its
spreads ({\it middle}) and a dual of the Petersen graph, or a
2-spread ({\it right}).}
\end{figure}

To complete this interesting algebraic geometrical picture of
two-qubits there remains to be introduced one more important
geometrical object. The attentive reader might have noticed that
we have already employed two different kinds of the projective
lines defined over rings of order four and characteristic two,
viz. the line defined over the field $GF(4)$ and that defined over
the direct product ring $GF(2) \times GF(2)$; the former was seen
to answer to an {\it ovoid} of the generalized quadrangle, alias
to a set of five mutually non-commuting operators (Fig.\,1, {\it
right -- top}), while the latter corresponds to a {\it grid} of
nine points on six lines,\footnote{Also known as the {\it slim}
generalized quadrangle of order ($2, 1$); see, e.\,g., Refs.\,18
and 20. In fact, {\it both} the configurations depicted in Fig.\,1, {\it left}, 
are slim generalized quadrangles, one being the {\it dual} of the other.} 
alias to a Mermin's square of operators (Fig.\,1, {\it
left -- bottom}). There, however, exists {\it one more}
associative ring with unity of order four and characteristic two,
namely the (local) factor ring of polynomials $GF(2)[x]/\langle
x^{2}\rangle$.$^{12-14}$ This ring is also a subring of
$M_{2}(GF(2))$, so the corresponding projective line is expected
to play a role in our model, too. And this is indeed the case. As
demonstrated, for example, in Ref.\,6 (Table 3), the projective
line $P_{1}(GF(2)[x]/\langle x^{2} \rangle)$ features six points
any of which is neighbor to one and distant to the remaining four
points, comprising thus three pairs of neighbors. In the set of
Pauli operators this configuration is present as the sextuple of
operators commuting with a given operator; taking the latter to
be, e.\,g.,  $C_{13}$, the six operators in question, as readily
discerned from Table 2, are $\{$$C_{4}$,  $C_{5}$; $C_{7}$,
$C_{10}$; $C_{14}$, $C_{15}$$\}$, which indeed form three pairs of
commuting members (these pairs being separated from each other by
a semicolon). In the generalized quadrangle any such configuration
resides as the sextuple of points collinear with a given point. A
deeper understanding and a fuller appreciation of this observation
is acquired after introducing the concept of a {\it geometric
hyperplane}. A geometric hyperplane $H$ of a finite geometry is a
set of points such that every line of the geometry either contains
exactly one point of  $H$, or is completely contained in
$H$.$^{20,\,22}$ It is easy to verify that for the generalized
quadrangle of order two $H$ is of one of the following {\it three}
kinds:$^{22}$ 1) $H_{{\rm ov}}$, an ovoid (there are six such
hyperplanes); 2) $H_{{\rm cl}}(X)$, a set of points collinear with
a given point $X$, the point itself {\it in}clusive (there are 15
such hyperplanes); and 3) $H_{{\rm gr}}$, a grid as defined  above
(there are 10 such hyperplanes). One thus reveals a perfect parity
between the three kinds of the geometric hyperplanes of the
generalized quadrangle of order two and the three kinds of the
projective lines over the rings of four elements and
characteristic two embedded in our sub-configuration of
$P_{1}(M_{2}(GF(2)))$, giving rise to the three kinds of the
distinguished subsets of the Pauli operators of two-qubits, as
summarized in Table\,3.
\begin{table}[t]
\begin{center}
\caption{Three kinds of the distinguished subsets of the
generalized Pauli operators of two-qubits (TQ) viewed as the geometric
hyperplanes in the generalized quadrangle of order two (GQ) and/or as
the projective lines over the rings of order four and
characteristic two living in  the projective line $P_{1}(M_{2}(GF(2)))$ (PL).}
\vspace*{0.4cm}
\begin{tabular}{llll}
\hline \hline
\vspace*{-.3cm} \\
GQ & $H_{{\rm ov}}$  &     $H_{{\rm cl}}(X)\setminus\{X\}$ &         $H_{{\rm gr}}$  \\
PL & $P_{1}(GF(4))$  &      $P_{1}(GF(2)[x]/\langle x^{2} \rangle)$  &   $P_{1}(GF(2) \times GF(2))$  \\
TQ & set of five mutually    & set of six operators  & nine operators of a \\
   & non-commuting operators  & commuting with a given one & Mermin's square\\
\vspace*{-.3cm} \\
\hline \hline
\end{tabular}
\end{center}
\end{table}
As a final note, it is worth mentioning that the generalized
quadrangle of order two also resides in $P_{1}(M_{2}(GF(2)))$ as
the projective line over the so-called Jordan system of {\it
symmetric} two-by-two matrices over $GF(2)$,$^{23}$ or,
equivalently, as a generic hyperplane section of the Klein quadric
in the 5-dimensional projective space over $GF(2)$.$^{22}$

We have demonstrated that the basic properties of a system of two interacting spin-1/2 particles are uniquely
embodied in the (sub)geometry of a particular projective line, found to be equivalent to the generalized
quadrangle of order two. As such systems are the simplest ones exhibiting
phenomena like quantum entanglement and quantum non-locality and play, therefore, a crucial role in numerous
applications like quantum cryptography, quantum coding, quantum cloning/teleportation and/or quantum computing
to mention the most salient ones, our discovery thus not only offers a principally new geometrically-underlined
insight into their intrinsic nature, but also gives their applications a wholly new perspective and opens up rather
unexpected vistas for an algebraic geometrical modelling of their higher-dimensional counterparts.$^{24,25}$

\large
\vspace*{.4cm}
\noindent
{\bf Acknowledgements}\\
\normalsize
The first author thanks Prof.\,Hans Havlicek (Vienna University of Technology) for bringing to our attention
the description of the generalized quadrangle of order two as the projective line over the Jordan system
of the ring in question.
This work was partially supported by the
Science and Technology Assistance Agency under the contract $\#$
APVT--51--012704, the VEGA project $\#$ 2/6070/26 (both from
Slovak Republic), the trans-national ECO-NET project $\#$
12651NJ ``Geometries Over Finite Rings and the Properties of
Mutually Unbiased Bases" (France) and by the project 1ET400400410 of
the Academy of Sciences of the Czech Republic.

\end{document}